\documentclass[aps,prb,groupedaddress,a4paper,preprint,endfloats]{revtex4}
\usepackage{bm,graphicx,graphics,amsmath,amssymb,bm,epsfig,color,pifont}
\usepackage{euscript,tabularx,textcomp}

\begin{document}
\bibliographystyle{apsrev}
 
\title{Magnetic resonance spectroscopy of perpendicularly magnetized\\
  permalloy multilayer disks}

\author{G.~de Loubens} \affiliation{Service de Physique de l'{\'E}tat
  Condens{\'e}, CEA Orme des Merisiers, F-91191 Gif-Sur-Yvette}
\author{V.~V.~Naletov} \thanks{Also at Physics Department, Kazan State
  University, Kazan 420008 Russia} \affiliation{Service de Physique de
  l'{\'E}tat Condens{\'e}, CEA Orme des Merisiers, F-91191 Gif-Sur-Yvette}
\author{M.~Viret} \affiliation{Service de Physique de l'{\'E}tat
  Condens{\'e}, CEA Orme des Merisiers, F-91191 Gif-Sur-Yvette}
\author{O.~Klein} \affiliation{Service
  de Physique de l'{\'E}tat Condens{\'e}, CEA Orme des Merisiers, F-91191
  Gif-Sur-Yvette}
\author{H.~Hurdequint} \affiliation{Laboratoire de Physique des
  Solides, Universit\'e Paris-Sud, F-91405 Orsay}
\author{J.~Ben~Youssef} \affiliation{Laboratoire de Magn\'etisme de
  Bretagne, 6 Av. Le Gorgeu, F-29285 Brest}
\author{F.~Boust} \affiliation{ONERA, 29 avenue de la Division
  Leclerc, F-92322 Ch\^atillon}
\author{N.~Vukadinovic} \affiliation{Dassault Aviation, DGT/DTIAE, 78
  quai Marcel Dassault, F-92552 Saint-Cloud}

\date{\today}

\begin{abstract}
 Using a Magnetic Resonance Force Microscope, we compare the ferromagnetic resonance spectra of individual micron-size disks with identical diameter, 1~$\mu$m, but different layer structures. For a disk composed of a single 43.3~nm thick permalloy (Py) layer, the lowest energy mode in the perpendicular configuration is the uniform precession. The higher energy modes are standing spin-waves confined along the diameter of the disk. For a Cu(30)/Py(100)/Cu(30)~nm multilayer structure, it has been interpreted that the lowest energy mode becomes a precession localized at the Cu/Py interfaces. When the multilayer is changed to Py(100)/Cu(10)/Py(10)~nm, this localized mode of the thick layer becomes coupled to the precession of the thin layer.
\end{abstract}

\pacs{ {76.50.+g}{Ferromagnetic, antiferromagnetic, and ferrimagnetic
  resonances} }

\maketitle

The dynamical properties of micron sized ferromagnetic structures
provides new challenges and opportunities for novel magnetoelectronics
devices\cite{prinz:99}. A recent focus is on the study of the exact
nature of the normal modes in these small
structures\cite{hillebrands:03}. Experimental works in the
quasi-saturated\cite{hiebert:97, bailleul:06} and closure domain
structures states\cite{perzlmaier:05, park:03} have been recently
reported. As for numerical simulations, the interplay between the
short-distance exchange and long-range dipolar interactions makes
calculations \cite{boust:04, mcmichael:05} challenging. Moreover, the
recently discovered spin diffusion mechanism\cite{hurdequint:06,
  tserkovnyak:05} makes the dynamical properties of such small
structures still unsorted.

Ferromagnetic resonance (FMR) has been for a long time the basic tool
to study the microwave susceptibility of magnetic
samples\cite{wigen:84}. FMR uses a well-defined selection rule, where
the excitation field preferentially couples to the most uniform mode
in the sample. However, the limited sensitivity of standard FMR
implies that it can only be performed on arrays of micron size disks,
which statistically averages the spectra of many disks and makes it
insensitive to individual differences. In this paper, using a
mechanical-FMR we report on a study of \textit{individual} micron size
disks. It is shown that different multilayer structures yield
different characteristics for the lowest energy mode.

\begin{figure}
 \includegraphics[width=6.0cm]{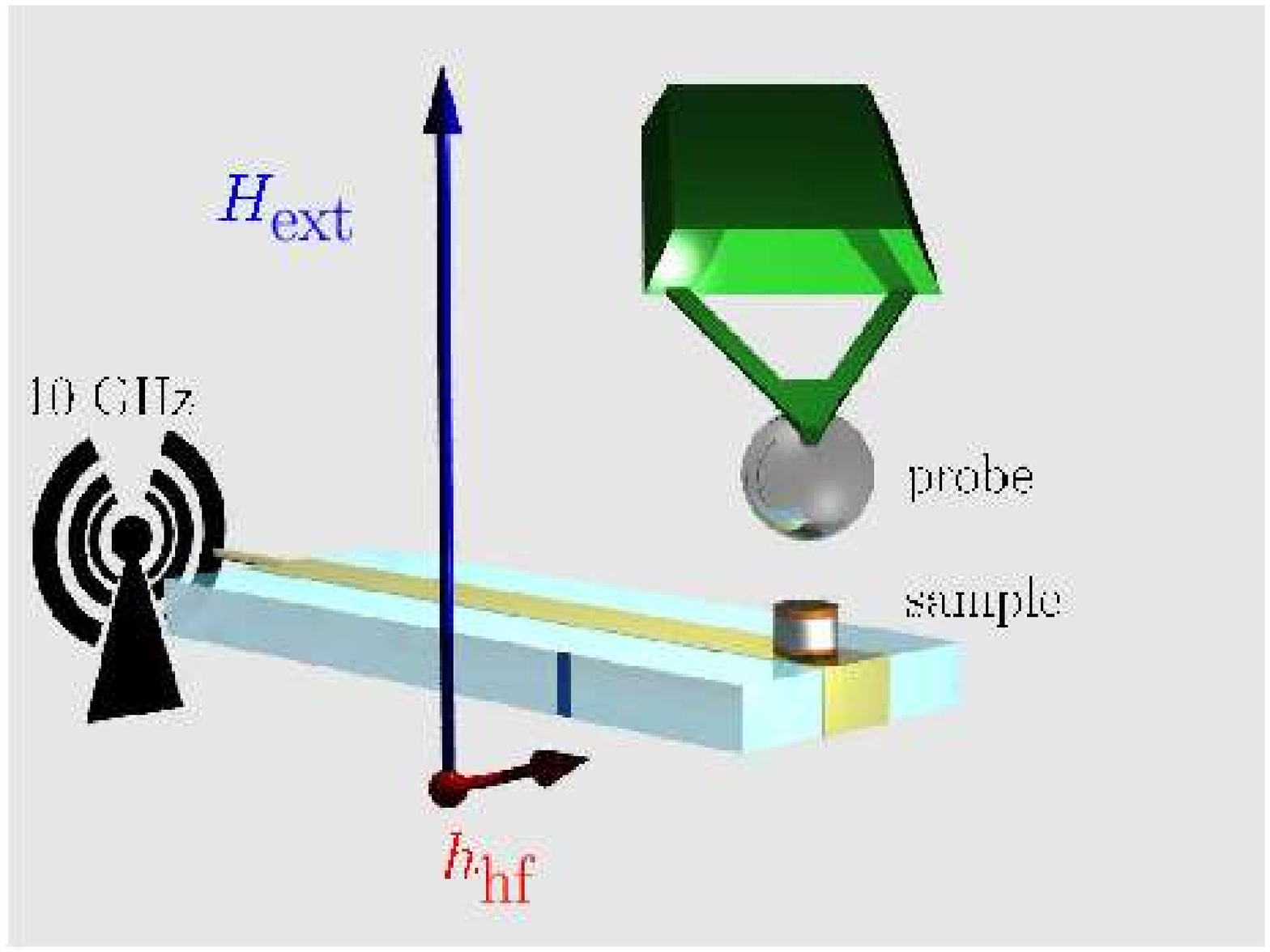}
 \caption{Schematic of the mechanical-FMR.}
 \label{fig1}
\end{figure}

To detect the FMR response of an individual microstructure, we exploit
the high sensitivity of mechanical-FMR\cite{zhang:96, charbois:02}, a
technique inspired by scanning probe techniques. As shown
schematically in Fig.~\ref{fig1}, the static part of the sample
magnetization is coupled through the dipolar interaction to a magnetic
sphere attached to the end of a very soft cantilever (spring constant
$5$~mN/m). Exciting the sample at a fixed frequency, the signal from
the sample is detected by measuring the cantilever motion as a
function of the perpendicular dc applied field, $H_{\text{ext}}$. A
force on the cantilever, proportional to the variation of longitudinal
magnetization (component along the precession axis), is associated
with the resonance\cite{naletov:03}. In the experiment reported here,
the sphere is an alloy whose principle constituents are (80 wt\%) Co
and (10 wt\%) Fe. It has a diameter of $4~\mu$m and its magnetic moment
is $(5\pm0.5) 10^{-9}$~emu.  The center of the sphere is positioned
above the center of the disk $(2.5\pm0.1)~\mu$m away from the sample
surface. In this geometry, the perturbation induced by the stray field
of the sphere on the FMR spectrum solely corresponds to a perturbation
shift of $-300$~Oe of the whole spectrum\cite{charbois:02}. To enhance
the sensitivity by the quality factor of the mechanical resonator
($Q=4500$), we measure with a lock-in the response of the cantilever
to a source (amplitude) modulation. The modulation frequency is then
set at the resonance frequency of the cantilever, which lies well
below all the relaxation rates in the spin system.

The excitation antenna, which generates the microwave field, is a
50$~\Omega$ microstrip circuit shorted at its extremity, a design allowing
studies in a broad frequency range. The microstrip is a Ti/Au(150~nm)
line deposited on a sapphire substrate.  The disks studied in the
present work have all the same diameter, 1~$\mu$m. They are patterned
by electron-beam lithography and ion-milling techniques out of
different multilayer thin films and are positioned at a magnetic field
anti-node of the microstrip.

\begin{figure}
 \includegraphics[width=8.0cm]{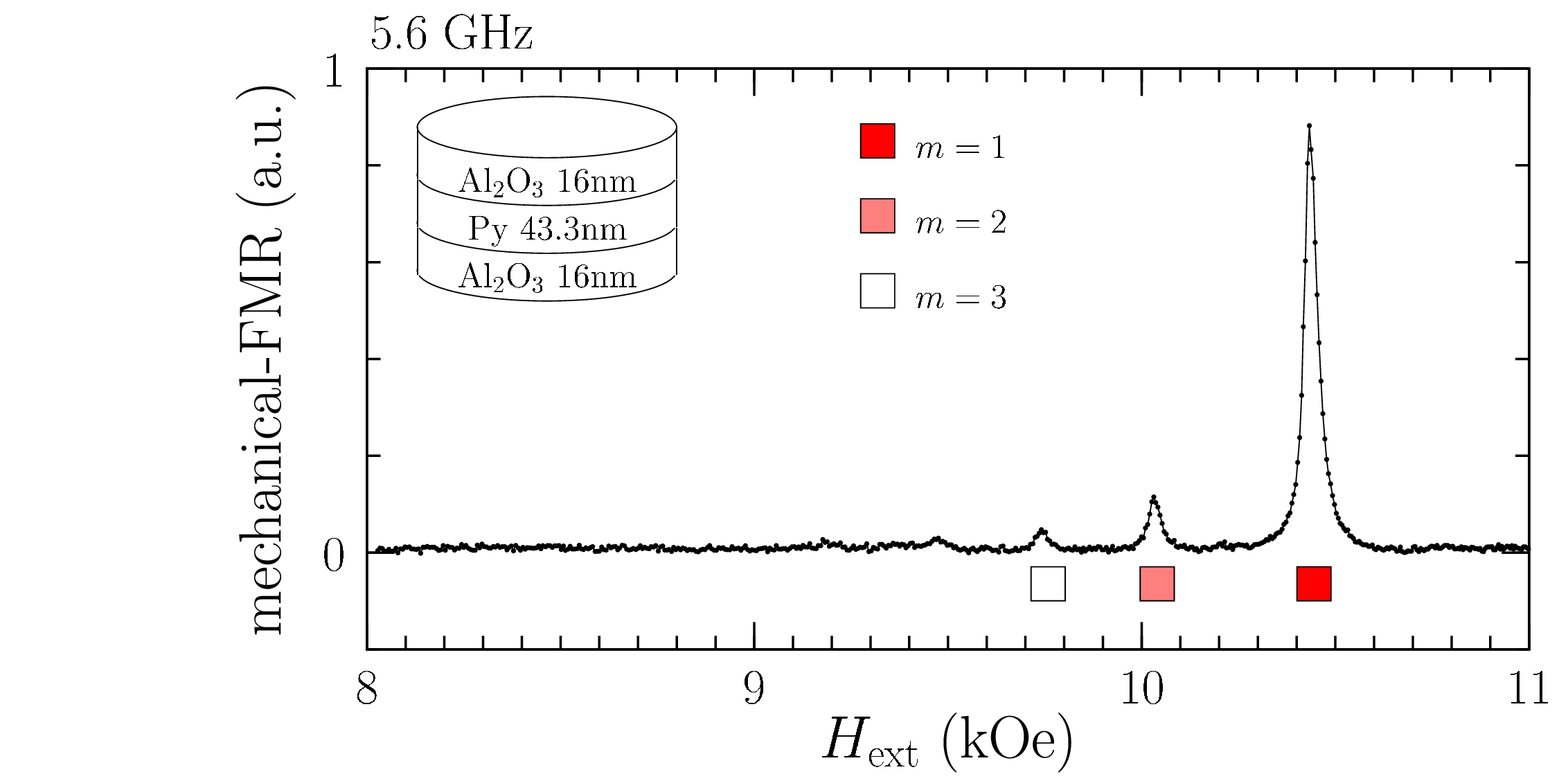}
 \caption{Mechanical-FMR spectrum of a single $t=43.3$~nm thick Py
  layer disk ($\phi=1~\mu$m) at 5.6GHz. The lowest energy mode
  corresponds to the uniform precession. The experimental positions
  of the magnetostatic modes are shown as squares.}
 \label{fig2}
\end{figure}

\begin{figure}
 \includegraphics[width=8.0cm]{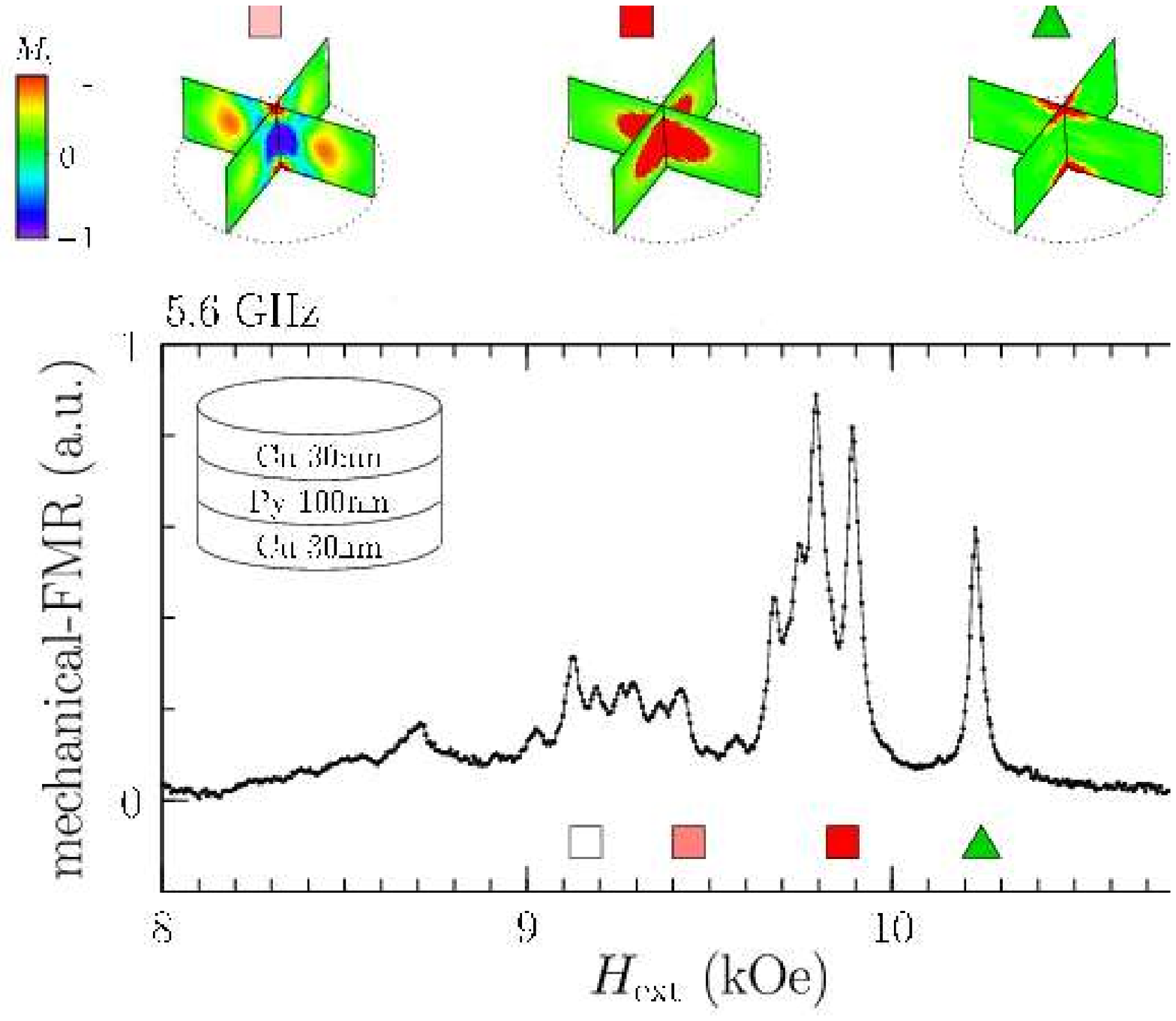}
 \caption{Mechanical-FMR spectrum of a single $t=100$~nm thick Py
  layer disk ($\phi=1~\mu$m) at 5.6GHz. The lowest energy mode
  is interpreted as a mode localized at the surface of the disk,
  shown as a triangle. Squares denote the uniform and magnetostatic 
  modes. The top shows the transverse dynamics of the main modes 
  represented in a color code, calculated using a micromagnetic simulation.}
 \label{fig3}
\end{figure}

\begin{figure}
 \includegraphics[width=8.0cm]{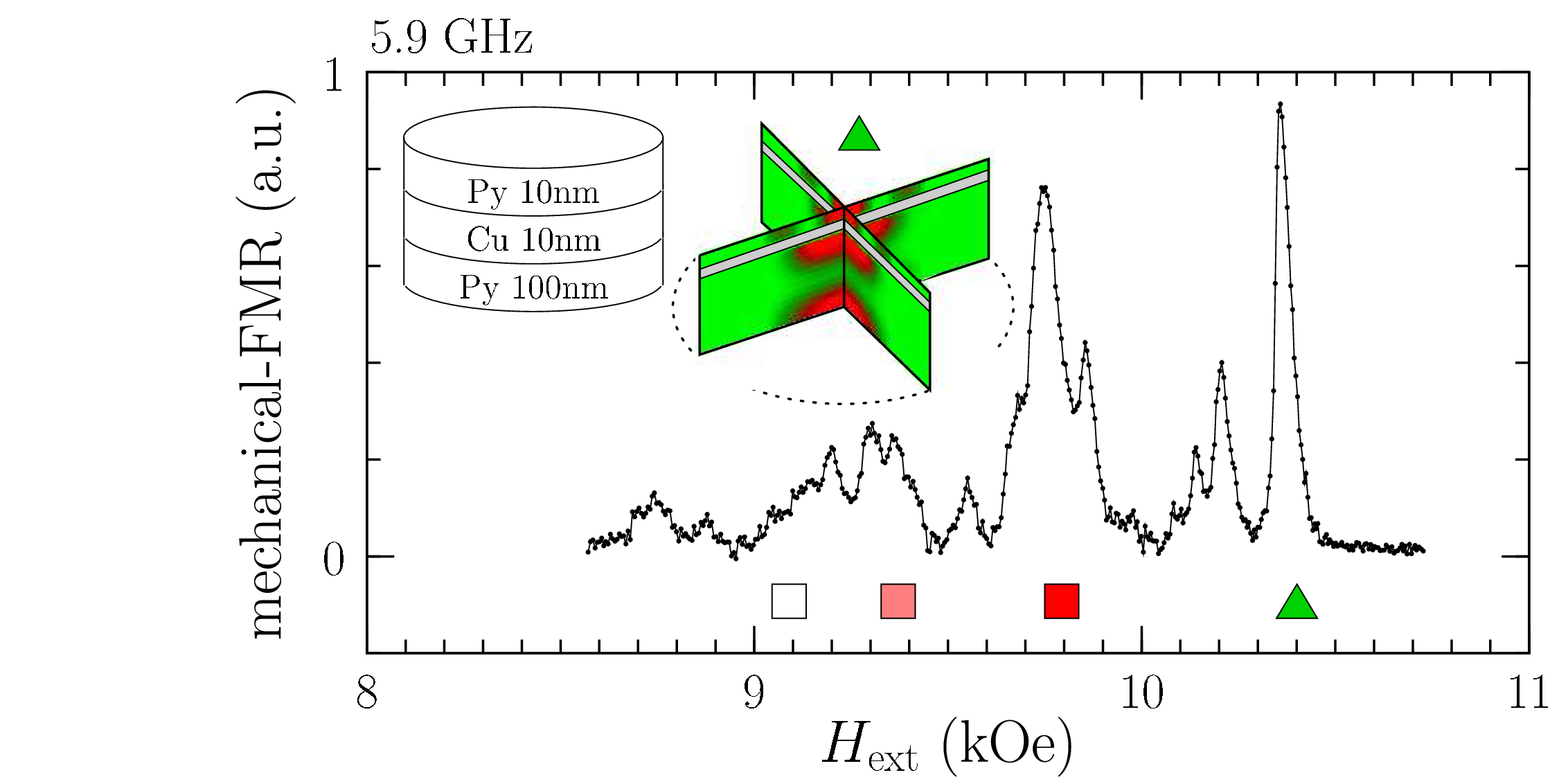}
 \caption{Mechanical-FMR spectrum of a double layer Py disk ($\phi=1~\mu$m)
   with thicknesses of 10~nm and 100~nm at 5.9~GHz. The spacer is a
   10~nm thick Cu layer. The lowest energy mode is associated to a
   coupled precession between the localized mode of the thick layer
   and the dynamics of the thin layer. The inset is a result of a
   micromagnetic simulation.}
 \label{fig4}
\end{figure}

Fig.~\ref{fig2} shows the 5.6~GHz mechanical-FMR spectrum of an
individual $\phi=1~\mu$m disk composed of a single $t=43.3$~nm thick
permalloy (Py) layer sandwiched by two insulating layers of
Al$_2$O$_3$ (16~nm).  The spectrum displayed on Fig.~\ref{fig2}, as
well as the ones of Fig.~\ref{fig3} and \ref{fig4}, has been recorded
at a stabilized temperature $T=280$~ K and in the linear regime, where
the peak amplitude remains proportional to the excitation power
(\textit{i.e.}  precession angles limited to 1\textdegree). The most intense
peak located at the highest field (lowest energy), $H_u\approx10.5$~kOe, is
the uniform mode.  Smaller amplitude modes at lower field (higher
energy) correspond to standing spin-wave (or magnetostatic modes) with
an increasing order $m$ along the radial direction\cite{kakazei:04}.
Note that the linewidth (31~Oe) of the peaks is among the smallest
reported at this frequency for Py. Such a regular spectrum has been
also measured by mechanical-FMR on an individual ferrite microdisk
\cite{charbois:02}.  It has also been observed using a cavity-FMR
technique by Kakazei \textit{et al.}\cite{kakazei:04} on an
\textit{array} of Py disks of same diameter and similar thickness.
The position of the magnetostatic modes can be calculated
analytically\cite{kakazei:04}. Using the physical parameters
characterizing our Py film deduced from X-band cavity-FMR performed on
the extended film\cite{hurdequint:02} from which the disk was
patterned out, we find that we have to introduce a small misalignment
(5\textdegree) with respect to the perpendicular direction to fit our data,
without any other fitting parameter.

Fig.\ref{fig3} shows the 5.6~GHz mechanical-FMR spectrum obtained for
a thicker Py layer ($t=100$nm) sandwiched by two Cu layers of 30nm.
There are several important differences with the previous spectrum:

$i$- The peaks of largest amplitude occur around $H_u\approx9.8$~kOe (red
square symbol). They should be associated to the largest volume of
precession (uniform mode). Lower value of the resonance field $H_u$ is
consistent with our previous measurement, since $H_u$ decreases with
increasing thickness. This is the consequence of the decrease of the
dipolar field as the aspect ratio ($t/\phi$) increases. The expected
shift, $\approx -800$~Oe, is close to the experimental one. The uniform mode
is followed by small peaks at lower fields, whose amplitudes are
consistent with their magnetostatic nature.
   
$ii$- While there is a single mode at the resonance field of the
uniform precession in the single layer (Fig.\ref{fig2}), the main mode
(red square) is split into at least two narrow peaks of 39~Oe
linewidth for the (Py/Cu) layers.  Such an observed spectrum points to
the presence of significant magnetic inhomogeneities in the film of
this disk sample.  We indicate that this (Py/Cu) layers have been
deposited on the microstrip Au layer. As reported in
ref\cite{loubens:06}, a detailed study of the FMR spectrum of the
control film (deposited on a Si substrate) has clearly revealed the
presence of such inhomogeneities: gradient of magnetization along the
thickness and roughness at the Cu/Py interfaces.

$iii$- The other striking feature is the appearance of a new mode at
$H_l\approx10.3$~kOe (green triangle) below the energy of the uniform
mode.  This new peak is interpreted\cite{loubens:06} as a precession
localized at the top and bottom Cu/Py interface near the disk center,
in the region of minimum internal field. Indeed, spins excited in the
minima region lead to a mode lower in energy than the core precession
because the gain in demagnetizing energy exceeds the cost in exchange
energy\cite{jorzick:02}. This explanation is consistent with the
absence of the localized mode in the spectrum of Fig.\ref{fig2}.

A cartography of the transverse component, obtained with a
micromagnetic code\cite{boust:04} in which the magnetic properties of
our Py have been used and where the two Cu layers have been neglected,
is displayed using a color code for the first three main modes. The
largest peak corresponds to the so-called ``uniform'' mode where all
the spins precess in phase in the bulk of the sample. The mode at
lower field (pink square) is the magnetostatic mode $m=2$ identified
in Fig.\ref{fig2} with a nodal circle in the disk (a locus where the
magnetization stays still along the equilibrium axis) and a
cylindrical regions precessing in opposition of phase with the core.
The same visualization of the lowest energy mode shows a precession
localized at the minima of the internal field. 

Fig.~\ref{fig4} shows the 5.9~GHz mechanical-FMR spectrum for a double
layer structure which consists of Py(100)/Cu(10)/Py(10)~nm. It is
directly deposited on the microstrip gold layer, and is caped by a
10~nm gold layer. This spectrum bears some strong resemblance with the
one observed in Fig.\ref{fig3}. This is expected since the main
features should be dominated by the dynamics of the thick layer. In
particular, we recognize the same group of large peaks at
$H_u=9.8$~kOe (red square), suggesting that this mode also corresponds
to the uniform precession of the thick layer. However, the largest
peak is now the lowest energy mode at $H_l=10.4$~kOe instead of the
uniform mode, and a new group of peaks appears between those two
modes. These two new features are likely related to the presence of
the thin Py layer.  Numerical simulations where the Cu spacer has been
replaced by a vacuum layer have been performed on this structure. The
simulated spectrum (not shown) gives the uniform mode of the thick
layer as well as a higher field and higher amplitude mode. The latter
corresponds to a localized precession of the thick layer coupled here
through the dipolar interaction to the magnetization dynamics of the
thin layer, shown in the inset of Fig.~\ref{fig4}.  One can understand
that this mode may have a larger amplitude than the uniform precession
since the top and bottom interfaces of the thick layer \textit{and}
the thin layer are now precessing at $H_l$. Physical process mediated
by the normal metal layer, which are neglected in the simulation (e.g.
spin diffusion), should enhance this coupling mechanism. We emphasize
that the simulation does not explain the new group of peaks observed
between the coupled localized and the uniform modes.  However, this
high field feature should be related to the dynamics inside the thin
layer (probably coupled to the dynamics of the thick layer), where the
internal field inhomogeneity and the quantization of the spin-wave
modes can give rise to several localized modes\cite{bayer:05}.

In conclusion, we have shown that different types of multilayer
patterned into the same microstructure yield different FMR spectra.
Importantly, the nature of the lowest energy mode depends on the
thickness of the Py layer. It shifts from the uniform mode to a
localized mode at the surfaces as the thickness increases. The
influence of a thinner layer separated by a normal metal spacer has
also been measured. We emphasize that the multiplicity of peaks for
the two thicker disks reported above have been attributed to extrinsic
effects\cite{loubens:06} and does not alter the main conclusions of
this work. For a finer interpretation of the reported spectra, such
effects should be taken into account, as well as the intrinsic effects
associated to the diffusion of the microwave
magnetization\cite{hurdequint:06} inside the whole metallic layered
structure.

We are greatly indebted to A.N. Slavin, X. Waintal, and A.-L. Adenot
for their help and support. This research was partially supported by
the ANR grant PNANO06-0235.


\begin{thebibliography}{20}
\expandafter\ifx\csname natexlab\endcsname\relax\def\natexlab#1{#1}\fi
\expandafter\ifx\csname bibnamefont\endcsname\relax
 \def\bibnamefont#1{#1}\fi
\expandafter\ifx\csname bibfnamefont\endcsname\relax
 \def\bibfnamefont#1{#1}\fi
\expandafter\ifx\csname citenamefont\endcsname\relax
 \def\citenamefont#1{#1}\fi
\expandafter\ifx\csname url\endcsname\relax
 \def\url#1{\texttt{#1}}\fi
\expandafter\ifx\csname urlprefix\endcsname\relax\def\urlprefix{URL }\fi
\providecommand{\bibinfo}[2]{#2}
\providecommand{\eprint}[2][]{\url{#2}}

\bibitem[{\citenamefont{Prinz}(1999)}]{prinz:99}
\bibinfo{author}{\bibfnamefont{G.~A.} \bibnamefont{Prinz}},
 \bibinfo{journal}{J. {M}agn. {M}agn. {M}ater.}
 \textbf{\bibinfo{volume}{200}}, \bibinfo{pages}{57} (\bibinfo{year}{1999}).

\bibitem[{\citenamefont{Hillebrands and Ounadjela}(1996)}]{hillebrands:03}
\bibinfo{author}{\bibfnamefont{B.}~\bibnamefont{Hillebrands}} \bibnamefont{and}
 \bibinfo{author}{\bibfnamefont{K.~E.} \bibnamefont{Ounadjela}},
 \emph{\bibinfo{title}{Spin Dynamics in Confined Magnetic Structures}},
 vol.~\bibinfo{volume}{2} (\bibinfo{publisher}{Springer, Berlin},
 \bibinfo{year}{1996}).

\bibitem[{\citenamefont{Hiebert et~al.}(1997)}]{hiebert:97}
\bibinfo{author}{\bibfnamefont{W.~K.} \bibnamefont{Hiebert}}
 \bibnamefont{et~al.}, \bibinfo{journal}{Phys. {R}ev. {L}ett.}
 \textbf{\bibinfo{volume}{79}}, \bibinfo{pages}{1134} (\bibinfo{year}{1997}).

\bibitem[{\citenamefont{Bailleul et~al.}(2006)}]{bailleul:06}
\bibinfo{author}{\bibfnamefont{M.}~\bibnamefont{Bailleul}}
 \bibnamefont{et~al.}, \bibinfo{journal}{Phys. {R}ev. {B}}
 \textbf{\bibinfo{volume}{73}}, \bibinfo{pages}{104424}
 (\bibinfo{year}{2006}).

\bibitem[{\citenamefont{Perzlmaier et~al.}(2005)}]{perzlmaier:05}
\bibinfo{author}{\bibfnamefont{K.}~\bibnamefont{Perzlmaier}}
 \bibnamefont{et~al.}, \bibinfo{journal}{Phys. {R}ev. {L}ett.}
 \textbf{\bibinfo{volume}{94}}, \bibinfo{pages}{057202}
 (\bibinfo{year}{2005}).

\bibitem[{\citenamefont{Park et~al.}(2003)}]{park:03}
\bibinfo{author}{\bibfnamefont{J.~P.} \bibnamefont{Park}} \bibnamefont{et~al.},
 \bibinfo{journal}{{P}hys. {R}ev. {B}} \textbf{\bibinfo{volume}{67}},
 \bibinfo{pages}{020403(R)} (\bibinfo{year}{2003}).

\bibitem[{\citenamefont{Boust and Vukadinovic}(2004)}]{boust:04}
\bibinfo{author}{\bibfnamefont{F.}~\bibnamefont{Boust}} \bibnamefont{and}
 \bibinfo{author}{\bibfnamefont{N.}~\bibnamefont{Vukadinovic}},
 \bibinfo{journal}{Phys. {R}ev. {B}} \textbf{\bibinfo{volume}{70}},
 \bibinfo{pages}{172408} (\bibinfo{year}{2004}).

\bibitem[{\citenamefont{McMichael and Stiles}(2005)}]{mcmichael:05}
\bibinfo{author}{\bibfnamefont{R.~D.} \bibnamefont{McMichael}}
 \bibnamefont{and} \bibinfo{author}{\bibfnamefont{M.~D.}
 \bibnamefont{Stiles}}, \bibinfo{journal}{J. {A}ppl. {P}hys.}
 \textbf{\bibinfo{volume}{97}}, \bibinfo{eid}{10J901}
 (pages~\bibinfo{numpages}{3}) (\bibinfo{year}{2005}).

\bibitem[{\citenamefont{Hurdequint}(2006)}]{hurdequint:06}
  \bibinfo{author}{\bibfnamefont{H.}~\bibnamefont{Hurdequint}},
  \bibinfo{year}{to be published in J. {M}agn. {M}agn. {M}ater. 2006}.

\bibitem[{\citenamefont{Tserkovnyak et~al.}(2005)}]{tserkovnyak:05}
\bibinfo{author}{\bibfnamefont{Y.}~\bibnamefont{Tserkovnyak}}
 \bibnamefont{et~al.}, \bibinfo{journal}{Rev. {M}od. {P}hys.}
 \textbf{\bibinfo{volume}{77}}, \bibinfo{pages}{1375} (\bibinfo{year}{2005}).

\bibitem[{\citenamefont{Wigen}(1984)}]{wigen:84}
\bibinfo{author}{\bibfnamefont{P.~E.} \bibnamefont{Wigen}},
 \bibinfo{journal}{Thin {S}olid {F}ilms} \textbf{\bibinfo{volume}{114}},
 \bibinfo{pages}{135} (\bibinfo{year}{1984}).

\bibitem[{\citenamefont{Zhang et~al.}(1996)}]{zhang:96}
\bibinfo{author}{\bibfnamefont{Z.}~\bibnamefont{Zhang}} \bibnamefont{et~al.},
 \bibinfo{journal}{Appl. {P}hys. {L}ett.} \textbf{\bibinfo{volume}{68}},
 \bibinfo{pages}{2005} (\bibinfo{year}{1996}).

\bibitem[{\citenamefont{Charbois et~al.}(2002)}]{charbois:02}
\bibinfo{author}{\bibfnamefont{V.}~\bibnamefont{Charbois}}
 \bibnamefont{et~al.}, \bibinfo{journal}{Appl. {P}hys. {L}ett.}
 \textbf{\bibinfo{volume}{80}}, \bibinfo{pages}{4795} (\bibinfo{year}{2002}).

\bibitem[{\citenamefont{Naletov et~al.}(2003)}]{naletov:03}
\bibinfo{author}{\bibfnamefont{V.~V.} \bibnamefont{Naletov}}
 \bibnamefont{et~al.}, \bibinfo{journal}{Appl. {P}hys. {L}ett.}
 \textbf{\bibinfo{volume}{83}}, \bibinfo{pages}{3132} (\bibinfo{year}{2003}).

\bibitem[{\citenamefont{Kakazei et~al.}(2004)}]{kakazei:04}
\bibinfo{author}{\bibfnamefont{G.~N.} \bibnamefont{Kakazei}}
 \bibnamefont{et~al.}, \bibinfo{journal}{Appl. {P}hys. {L}ett.}
 \textbf{\bibinfo{volume}{85}}, \bibinfo{pages}{443} (\bibinfo{year}{2004}).

\bibitem[{\citenamefont{Kalinikos and Slavin}(1986)}]{kalinikos:86}
\bibinfo{author}{\bibfnamefont{B.~A.} \bibnamefont{Kalinikos}}
 \bibnamefont{and} \bibinfo{author}{\bibfnamefont{A.~N.}
 \bibnamefont{Slavin}}, \bibinfo{journal}{J. {P}hys. {C}}
 \textbf{\bibinfo{volume}{19}}, \bibinfo{pages}{7013} (\bibinfo{year}{1986}).

\bibitem[{\citenamefont{Hurdequint}(2002)}]{hurdequint:02}
\bibinfo{author}{\bibfnamefont{H.}~\bibnamefont{Hurdequint}},
 \bibinfo{journal}{J. {M}agn. {M}agn. {M}ater.}
 \textbf{\bibinfo{volume}{242-245}}, \bibinfo{pages}{521}
 (\bibinfo{year}{2002}).

\bibitem[{\citenamefont{Loubens et~al.}(2006)}]{loubens:06}
\bibinfo{author}{\bibfnamefont{G.~} \bibnamefont{de~Loubens}}
 \bibnamefont{et~al.}, \bibinfo{journal}{cond-mat/0606245}
 (\bibinfo{year}{2006}).

\bibitem[{\citenamefont{Jorzick et~al.}(2002)}]{jorzick:02}
\bibinfo{author}{\bibfnamefont{J.}~\bibnamefont{Jorzick}} \bibnamefont{et~al.},
 \bibinfo{journal}{Phys. {R}ev. {L}ett.} \textbf{\bibinfo{volume}{88}},
 \bibinfo{pages}{47204} (\bibinfo{year}{2002}).

\bibitem[{\citenamefont{Bayer et~al.}(2005)}]{bayer:05}
\bibinfo{author}{\bibfnamefont{C.}~\bibnamefont{Bayer}} \bibnamefont{et~al.},
 \bibinfo{journal}{Phys. {R}ev. {B}} \textbf{\bibinfo{volume}{72}},
 \bibinfo{pages}{064427} (\bibinfo{year}{2005}).

\end{thebibliography}
\end{document}